\documentclass[twoside]{dis08}
\usepackage[latin1]{inputenc}
\usepackage[dvips]{graphicx,epsfig,color}
\usepackage{wrapfig,rotating}
\usepackage{amssymb,amsmath,array}

\def\alt{\:\raisebox{-0.5ex}{$\stackrel{\textstyle<}{\sim}$}\:}

\catcode`@=11
\newcount\@tempcntc
\def\@citex[#1]#2{\if@filesw\immediate\write\@auxout{\string\citation{#2}}\fi
  \@tempcnta\z@\@tempcntb\m@ne\def\@citea{}\@cite{\@for\@citeb:=#2\do
    {\@ifundefined
       {b@\@citeb}{\@citeo\@tempcntb\m@ne\@citea\def\@citea{,}{\bf ?}\@warning
       {Citation `\@citeb' on page \thepage \space undefined}}%
     {\setbox\z@\hbox{\global\@tempcntc0\csname b@\@citeb\endcsname\relax}%
     \ifnum\@tempcntc=\z@ \@citeo\@tempcntb\m@ne
       \@citea\def\@citea{,}\hbox{\csname b@\@citeb\endcsname}%
     \else
      \advance\@tempcntb\@ne
      \ifnum\@tempcntb=\@tempcntc
      \else\advance\@tempcntb\m@ne\@citeo
      \@tempcnta\@tempcntc\@tempcntb\@tempcntc\fi\fi}}\@citeo}{#1}}
\def\@citeo{\ifnum\@tempcnta>\@tempcntb\else\@citea\def\@citea{,}%
  \ifnum\@tempcnta=\@tempcntb\the\@tempcnta\else
   {\advance\@tempcnta\@ne\ifnum\@tempcnta=\@tempcntb \else \def\@citea{--}\fi
    \advance\@tempcnta\m@ne\the\@tempcnta\@citea\the\@tempcntb}\fi\fi}
\catcode`@=12

\begin{document}
\title{Inclusive production of heavy-flavored hadrons at NLO in the GM-VFNS}

\author{Bernd A. Kniehl
%
%
\vspace{.3cm}\\
%
II. Institut f\"ur Theoretische Physik, Universit\"at Hamburg,\\
Luruper Chaussee 149, 22761 Hamburg, Germany
%
}

\maketitle

\begin{abstract}
We summarize recent progress in the theoretical description of
heavy-flavored-hadron inclusive production at next-to-leading order in the
general-mass variable-flavor-number scheme.
Specifically, we discuss the influence of finite-mass effects on the
determination $D$-meson fragmentation functions from a global fit to
$e^+e^-$ annihilation data and on the transverse-momentum distribution of
$B$-meson hadroproduction.
We also demonstrate that the fixed-flavor-number scheme, implemented with
up-to-date parton density functions and strong-coupling constant, provides a
surprisingly good description of $B$-meson data from run~II at the Fermilab
Tevatron.
\end{abstract}

\section{Introduction}

The general-mass variable-flavor-number scheme (GM-VFNS) provides a rigorous
theoretical framework for the theoretical description of the inclusive
production of single heavy-flavored hadrons, combining the
fixed-flavor-number scheme (FFNS) and zero-mass variable-flavor-number scheme
(ZM-FVNS), which are valid in complementary kinematic regions, in a unified
approach that enjoys the virtues of both schemes and, at the same time, is
bare of their flaws.
Specifically, it resums large logarithms by the
Dokshitzer-Gribov-Lipatov-Altarelli-Parisi (DGLAP) evolution of
non-perturbative fragmentation functions (FFs), guarantees the universality
of the latter as in the ZM-VFNS, and simultaneously retains the mass-dependent
terms of the FFNS without additional theoretical assumptions.
It was elaborated at next-to-leading order (NLO) for photo- \cite{KS} and
hadroproduction \cite{KKSS,PRL}.
In this presentation, we report recent progress in the implementation of the
GM-VFNS at NLO.
In Sec.~\ref{sec:d}, we present mass-dependent FFs for $D$-mesons extracted
from global fits to $e^+e^-$ annihilation data \cite{Kneesch:2007ey}.
In Sec.~\ref{sec:b}, we compare with transverse-momentum ($p_T$) distributions
of $B$ mesons produced in run~II at the Tevatron \cite{Kniehl:2008zz}.
Our conclusions are summarized in Sec.~\ref{sec:conclusions}.

\boldmath
\section{$D$-meson fragmentation functions}
\label{sec:d}
\unboldmath

\begin{figure}[ht]
\begin{center}
\begin{tabular}{ccc}
\includegraphics[height=3.7cm]{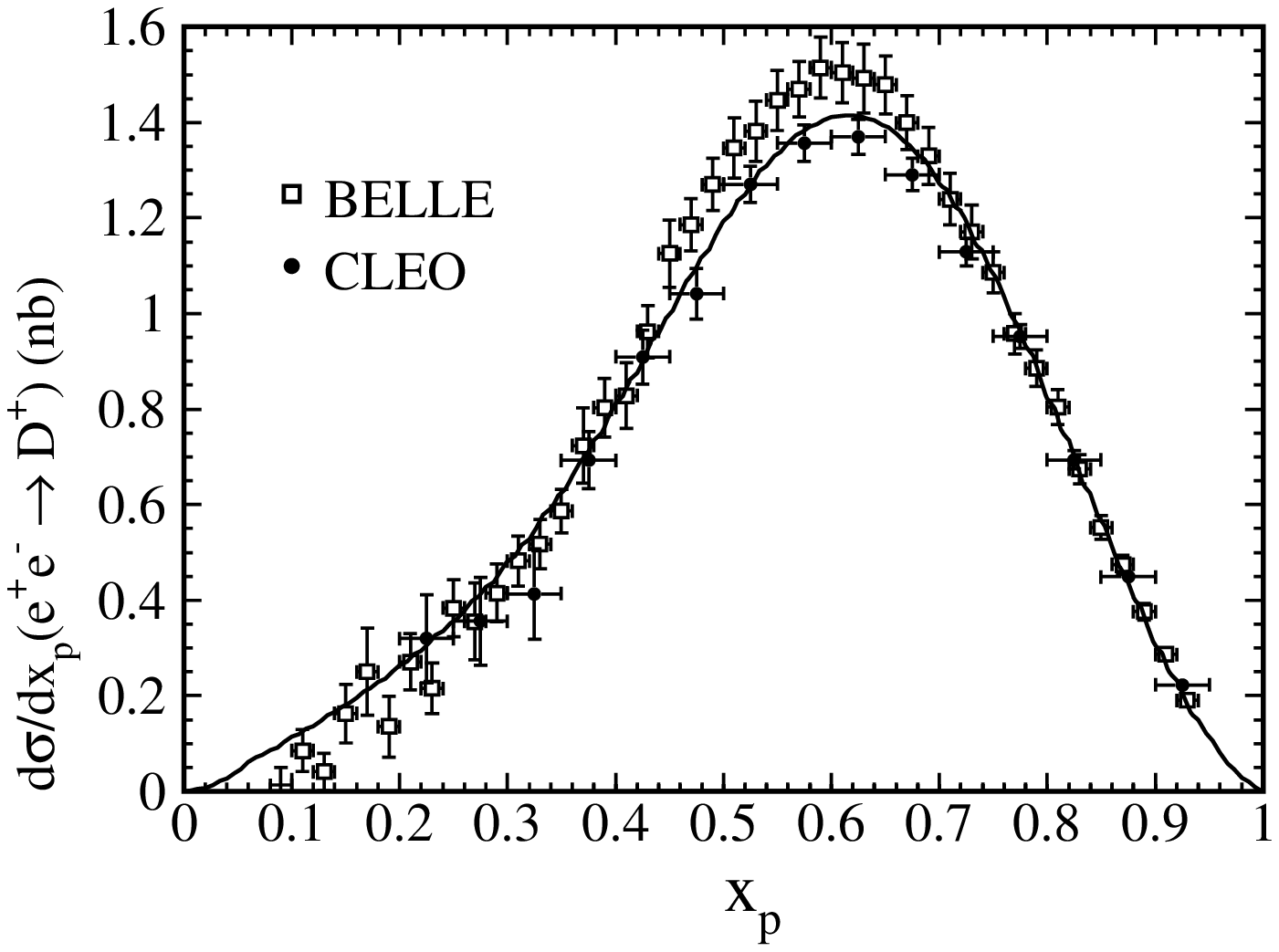} &
\includegraphics[height=3.7cm]{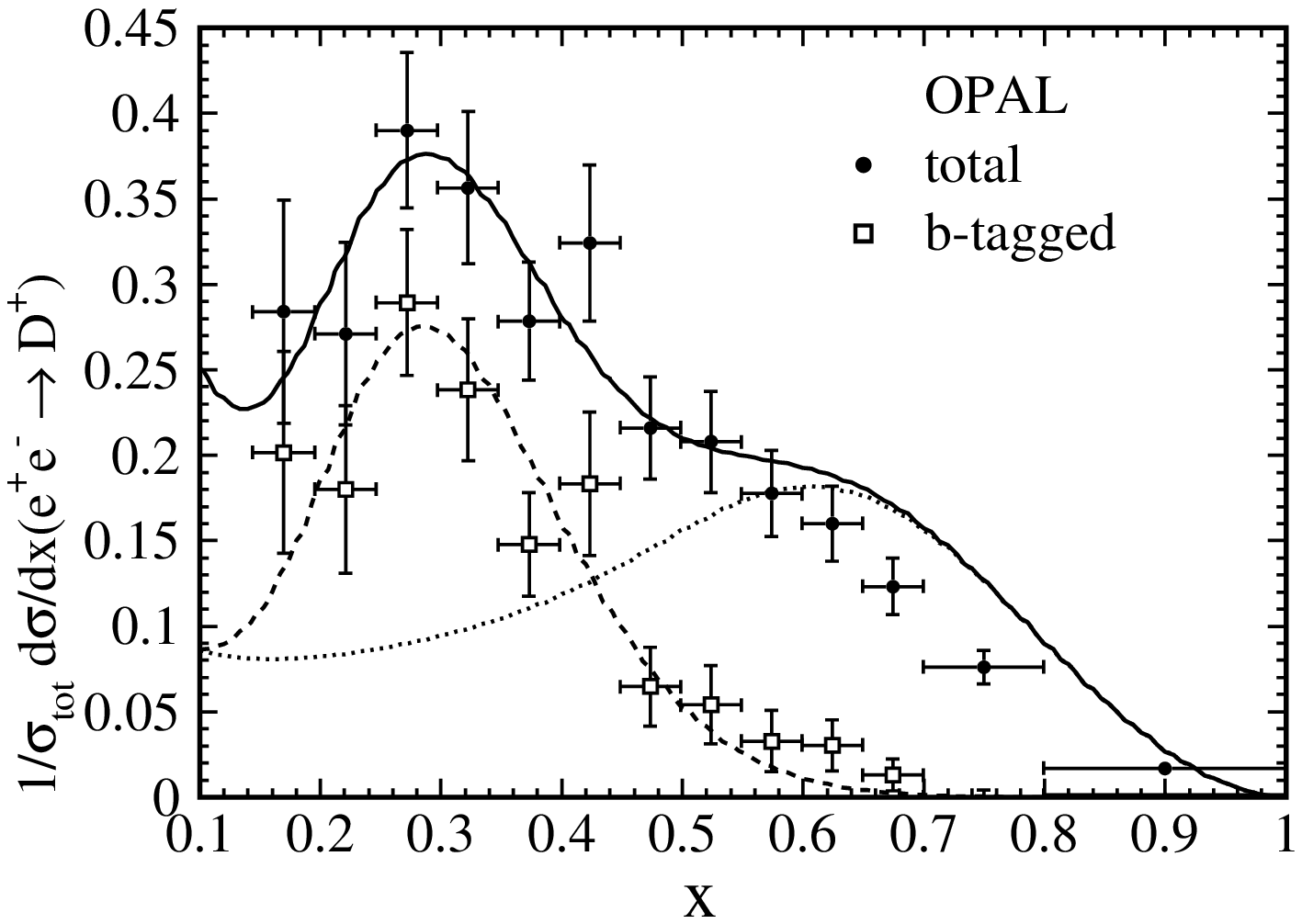} &
\includegraphics[height=3.7cm]{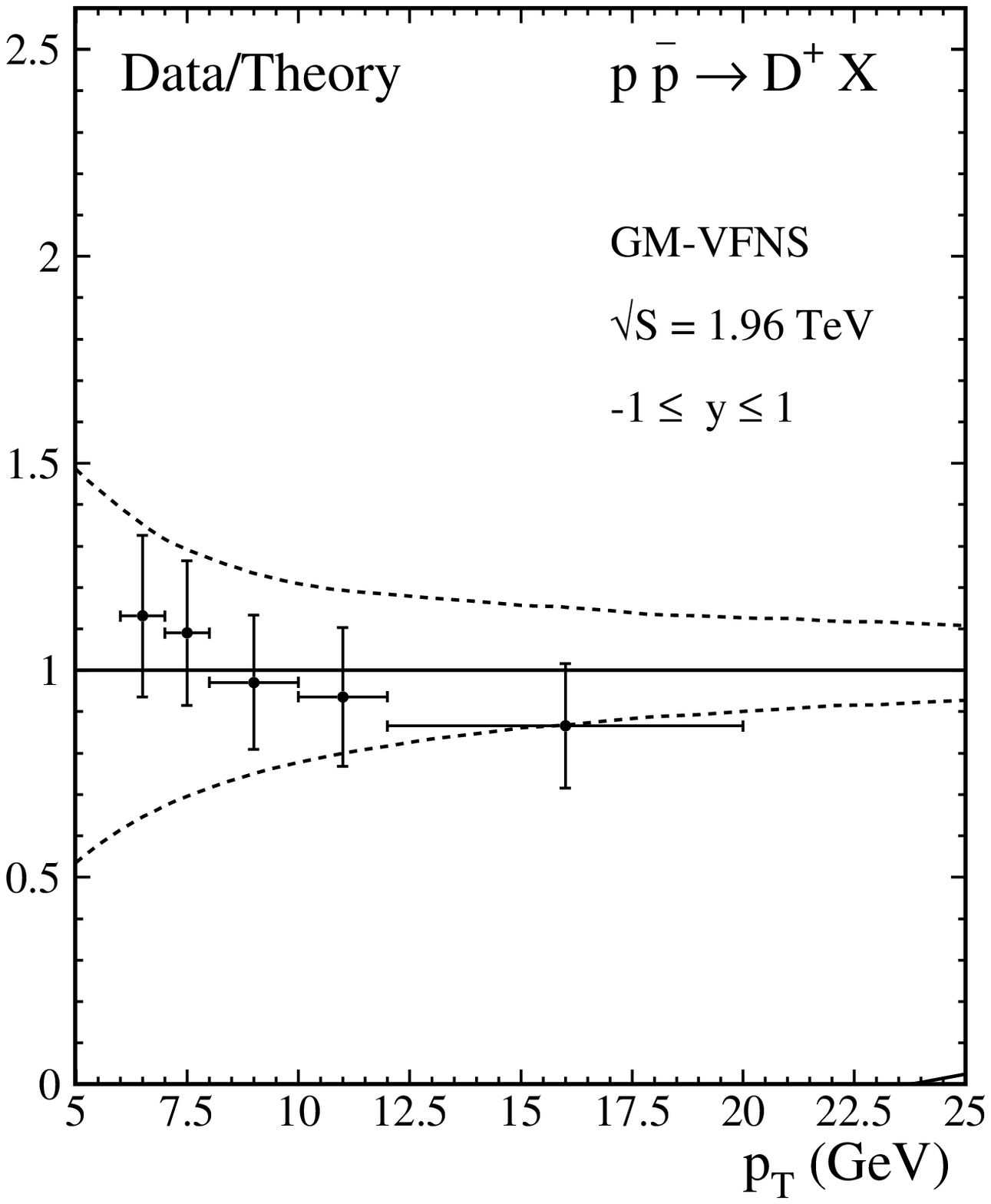}\\
(a) & (b) & (c) 
\end{tabular}
\end{center}
\vspace{-0.5cm}
\caption{Comparison of (a) Belle, CLEO, (b) ALEPH, OPAL \cite{ddata} (middle),
and (c) CDF~II data \cite{Acosta:2003ax} on $D^+$ mesons with global fit.
The dotted line in panel (b) refers to the $c$-quark-initiated contribution.}
\label{fig:d}
\end{figure}
In Ref.~\cite{Kneesch:2007ey}, we determined non-perturbative FFs for $D^0$,
$D^+$, and $D^{*+}$ mesons by fitting experimental data from the Belle, CLEO,
ALEPH, and OPAL Collaborations \cite{ddata}, taking dominant electroweak
corrections due to photonic initial-state radiation into account.
The fits for $D^0$, $D^+$, and $D^{*+}$ mesons using the Bowler ansatz
\cite{Bowler:1981sb} yielded $\chi^2/\mathrm{d.o.f.}=4.03$, 1.99, and 6.90,
respectively.
We assessed the significance of finite-mass effects through comparisons with a
similar analysis in the ZM-VFNS.
Under Belle and CLEO experimental conditions, charmed-hadron mass effects on
the phase space turned out to be appreciable, while charm-quark mass effects on
the partonic matrix elements are less important.
In Figs.~\ref{fig:d}(a) and (b), the scaled-momentum distributions from Belle
and CLEO and the normalized scaled-energy distributions from ALEPH and OPAL,
respectively, for $D^+$ mesons are compared to the global fits.
We found that the Belle and CLEO data tend to drive the average $x$ value of
the $c\to D$ FFs to larger values, which leads to a worse description of the
ALEPH and OPAL data.
Since the $b\to D$ FFs are only indirectly constrained by the Belle and CLEO
data, their form is only feebly affected by the inclusion of these data in the
fits.
Usage of these new FFs leads to an improved description of the CDF data
\cite{Acosta:2003ax} from run~II at the Tevatron, as may be seen by comparing
Fig.~\ref{fig:d}(c) with Fig.~2(b) of Ref.~\cite{PRL}.

\boldmath
\section{$B$-meson hadroproduction}
\label{sec:b}
\unboldmath

\begin{wrapfigure}{r}{0.3\columnwidth}
\centerline{\includegraphics[height=3.7cm]{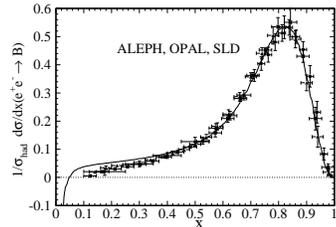}}
\vspace{-0.5cm}
\caption{Comparison of the ALEPH, OPAL, and SLD data \cite{bdata} on $B$
mesons with global fit.}\label{fig:b}
\end{wrapfigure}
\begin{figure}[ht]
\begin{center}
\begin{tabular}{ccc}
\includegraphics[height=4.5cm,viewport=40 4 420 471,clip]{%
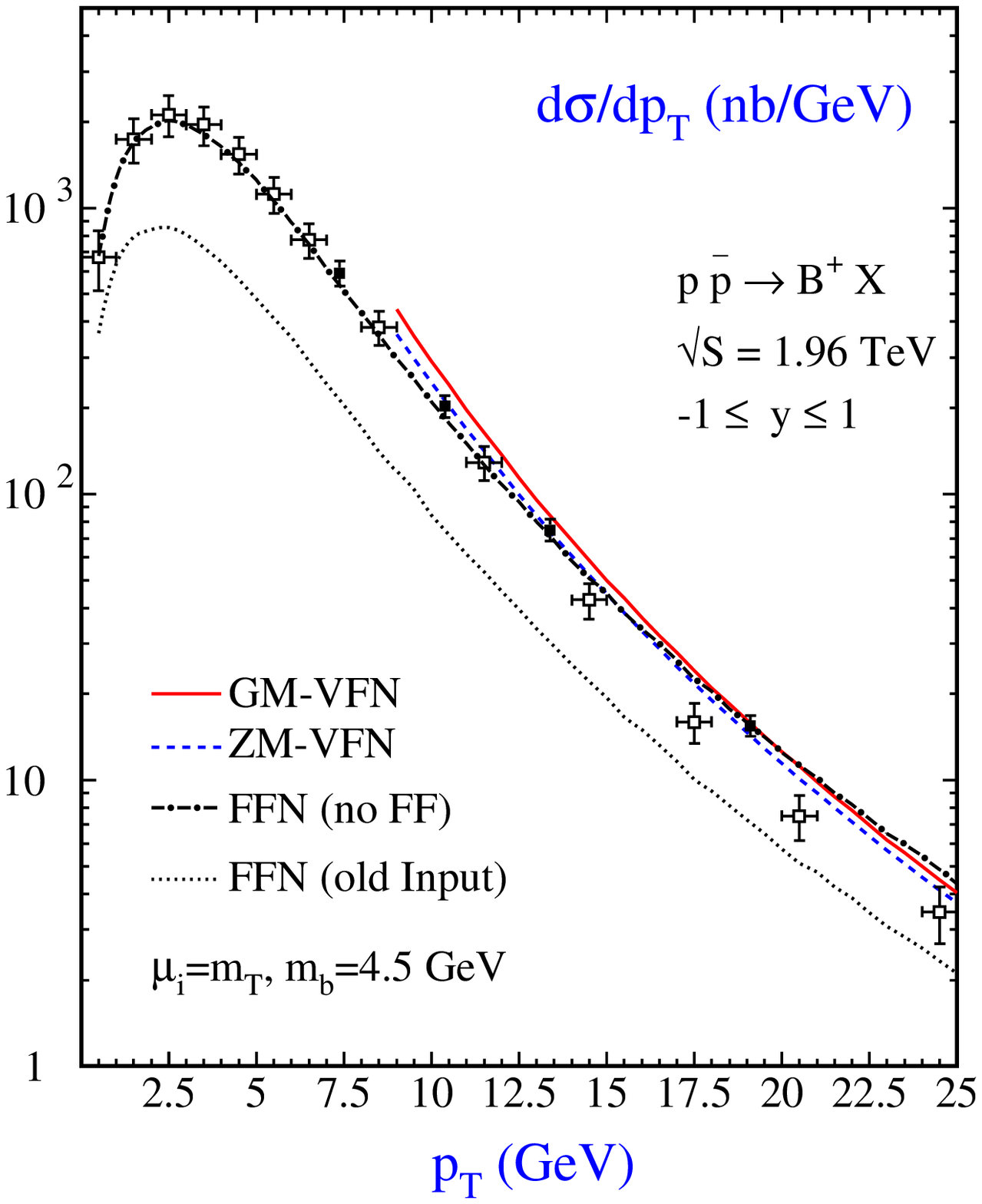} &
\includegraphics[height=4.5cm,viewport=12 0 518 494,clip]{%
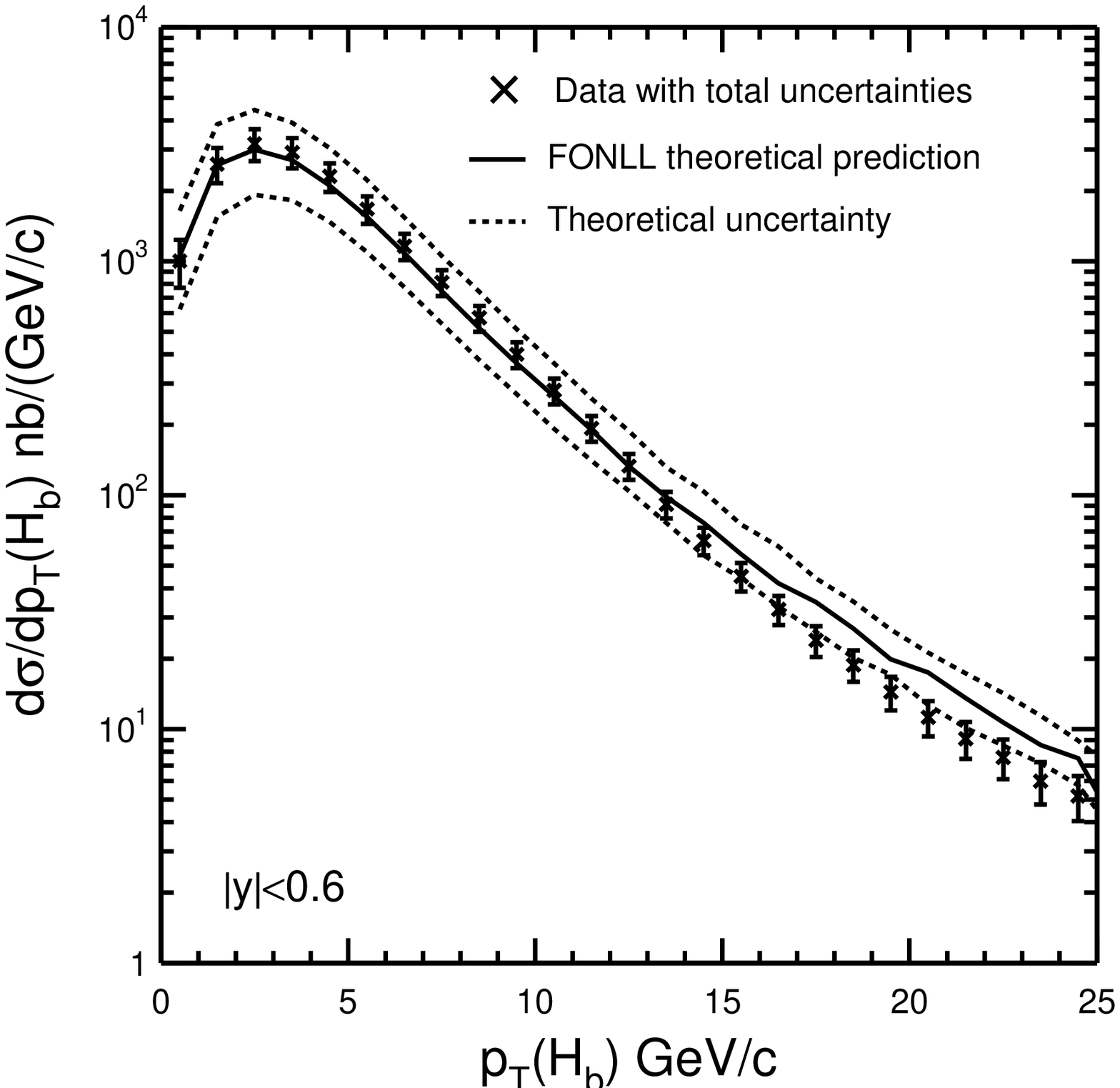} &
\includegraphics[height=4.5cm,viewport=14 18 283 293,clip]{%
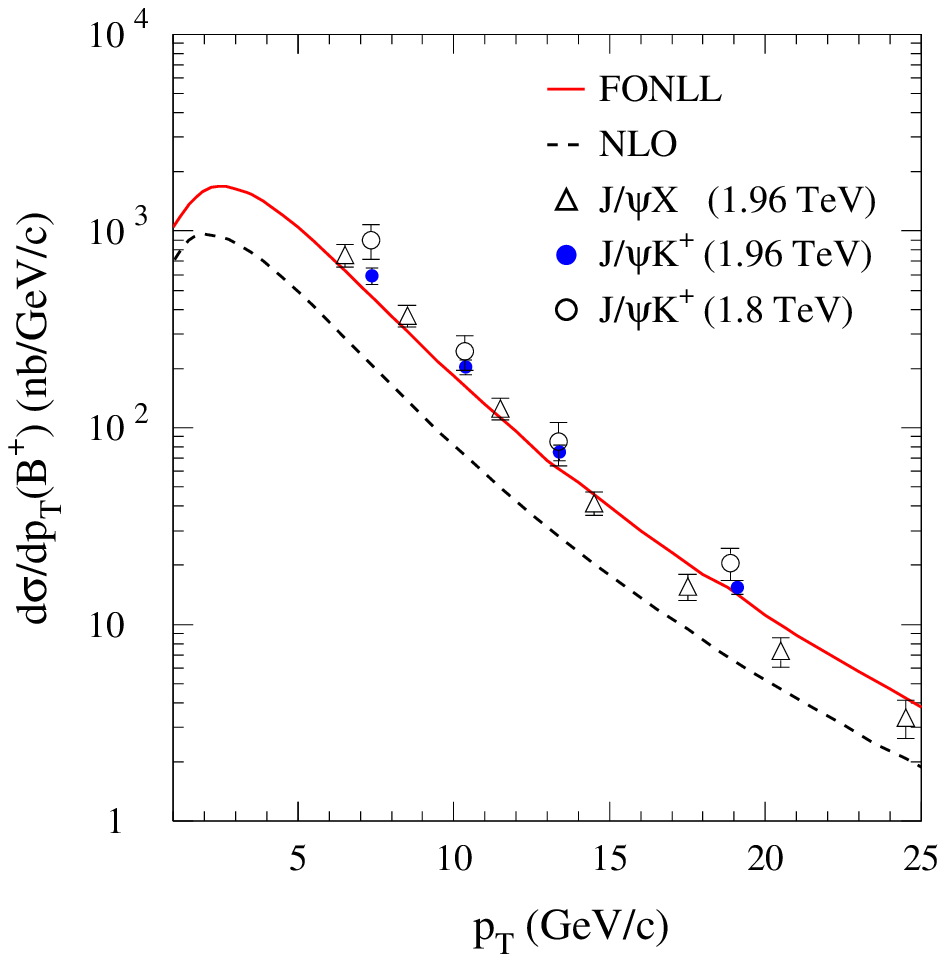} \\
(a) & (b) & (c) 
\end{tabular}
\end{center}
\vspace{-0.5cm}
\caption{Comparison of CDF~II data \cite{CDF1,CDF2} on $B$ mesons with
GM-VFNS, ZM-VFNS, FFNS, and FONLL predictions taken from Refs.\ (a)
\cite{Kniehl:2008zz}, (b) \cite{CDF1}, and (c) \cite{CDF2}.}
\label{fig:bcdf}
\end{figure}

In Ref.~\cite{Kniehl:2008zz}, we performed a comparative analysis of
$B$-meson hadroproduction in the ZM-VFNS and GM-VFNS. 
For this, we also updated the determination of $B$-meson FFs in the ZM-VFNS
\cite{BKK} by fitting to recent $e^+e^-$ data from ALEPH, OPAL, and SLD
\cite{bdata} and also adjusting the values of $m_b$ and the energy scale
$\mu_0$ where the DGLAP evolution starts to conform with modern PDF sets.
The fit using the Kartvelishvili-Likhoded ansatz \cite{Kartvelishvili:1985ac}
yielded $\chi^2/\mathrm{d.o.f.}=1.495$ (see Fig.~\ref{fig:b}).
We found that finite-$m_b$ effects moderately enhance the $p_T$ distribution;
the enhancement amounts to about 20\% at $p_T=2m_b$ and rapidly decreases with
increasing value of $p_T$, falling below 10\% at $p_T=4m_b$ (see
Fig.~\ref{fig:bcdf}a).
Such effects are thus comparable in size to the theoretical uncertainty due to
the freedom of choice in the setting of the renormalization and factorization
scales.
This finding contradicts earlier assertions~\cite{Cacciari:2003uh} that mass
corrections have a large size up to $p_T\approx20$~GeV and that {\it lack of
mass effects \cite{BKK} will therefore erroneously overestimate the production
rate at small $p_T$} in all respects.

In this connection, we also wish to point out that the statement made in
Ref.~\cite{Cacciari:2002xb} that {\it large logarithmic corrections in the
function $D(x,m^2)$ are simply discarded} in the approach of Ref.~\cite{BKK}
is misleading.
In fact, in the ZM-VFNS with non-perturbative FFs adopted in Ref.~\cite{BKK},
the Sudakov logarithms are fully included at NLO, namely both in the
coefficient functions and evolution kernels, and there is no room for large
logarithmic corrections in the ansatz for the heavy-quark FF at the initial
scale $\mu_0$, which represents non-perturbative input to be fitted to
experimental data.
Looking at Fig.~1 in Ref.~\cite{BKK}, we observe that the theoretical results
for $(1/\sigma_{\mathrm{had}})({\mathrm d}\sigma/{\mathrm d}x)(e^+e^-\to B+X)$
exhibit excellent perturbative stability and nicely agree with the OPAL data
\cite{OPAL} in the large-$x$ regime, indicating that Sudakov resummation is
dispensable in this scheme, in contrast to the
fixed-order-next-to-leading-logarithm (FONLL) scheme
\cite{Cacciari:2003uh,FONLL}, where the FFs are arranged to have perturbative
components.

\begin{wrapfigure}{r}{0.3\columnwidth}
\centerline{
\includegraphics[height=4.5cm,viewport=9 69 528 645,clip]{%
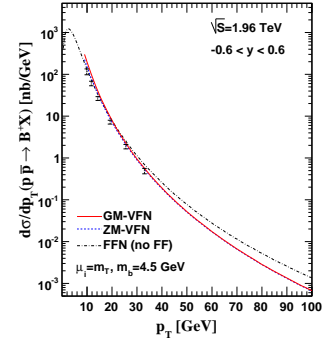}}
\vspace{-0.5cm}
\caption{Comparison of preliminary CDF~II data \cite{CDF3} on $B$ mesons
with GM-VFNS, ZM-VFNS, and FFNS predictions \cite{Kniehl:2008zz}.}
\label{fig:kkss}
\end{wrapfigure}
We must also caution the reader of the potential of comparisons of
experimental data with theoretical predictions in recent CDF~II publications
\cite{CDF1,CDF2} to be misinterpreted.
In Fig.~11 of Ref.~\cite{CDF1} (see Fig.~\ref{fig:bcdf}b), the variation of the
ad-hoc weight function, $G(m,p_T)=p_T^2/(p_T^2+c^2m^2)$ with $c=5$
\cite{Cacciari:2003uh,FONLL}, which has a crucial impact on the prediction in
the small-$p_T$ range by substantially suppressing its ZM-VFNS component, is
not included in the theoretical error.
In Fig.~11 of Ref.~\cite{CDF2} (see Fig.~\ref{fig:bcdf}c), the FFNS result,
labeled {\it NLO}, is evaluated with the obsolete MRSD0 proton PDFs
\cite{Martin:1992as}, revoked by their authors long ago, and a value of
$\alpha_s^{(5)}(m_z)$ falling short of the present world average \cite{pdg} by
3.3 standard deviations.
Unfortunately, this historical result is still serving as a benchmark
\cite{Happacher}.
Despite unresummed large logarithms and poorly implemented fragmentation, the
FFNS prediction, evaluated with up-to-date input, happens to almost coincide
with the GM-VFNS one in the range 15~GeV${}\alt p_T\alt25$~GeV.
It also nicely reproduces the peak exhibited about $p_T\approx2.5$~GeV by the
CDF~II data of Ref.~\cite{CDF1} (see Fig.~\ref{fig:bcdf}a).

In Fig.~\ref{fig:kkss}, preliminary CDF~II data \cite{CDF3}, which explore the
range 25~GeV${}<p_T<40$~GeV for the first time, are compared with NLO
predictions in the GM-VFNS, ZM-VFNS, and FFNS \cite{Kniehl:2008zz}.
In the large-$p_T$ limit, the GM-VFNS result steadily merges with the ZM-VFNS
one as per construction, while the FFNS breaks down due to unresummed large
logarithms.
The CDF~II data point in the bin 29~GeV${}<p_T<40$~GeV favors the GM-VFNS and
ZM-VFNS results, while it undershoots the FFNS result.

\section{Conclusions}
\label{sec:conclusions}

The GM-VFNS provides a rigorous theoretical framework for global analyses of
heavy-flavored-hadron inclusive production, retaining the full mass dependence
of the FFNS, preserving the scaling violations and universality of the FFs in
the ZM-VFNS, avoiding spurious $x\to1$ problems, and doing without ad-hoc
weight functions.
It has been elaborated at NLO for single production in $\gamma\gamma$,
$\gamma p$ \cite{KS}, $p\overline{p}$ \cite{KKSS,PRL,Kniehl:2008zz}, and
$e^+e^-$ collisions \cite{Kneesch:2007ey}.
More work is in progress.

\section*{Acknowledgments}

The author thanks T.~Kneesch, G.~Kramer, I.~Schienbein, and H.~Spiesberger for
the collaboration on the work presented here.
This work was supported in part by
DFG
Grant No.\ KN~365/7--1 and by 
BMBF Grant No.\ 05~HT6GUA.
 

\begin{footnotesize}

\end{footnotesize}


\begin{thebibliography}{99}

\bibitem{url} Slides: \\ 
\verb$http://indico.cern.ch/contributionDisplay.py?contribId=237&sessionId=14&confId=24657$

\bibitem{KS}G. Kramer and H. Spiesberger,
Eur.\ Phys.\ J.\ {\bf C22} 289 (2001);
{\bf C28} 495 (2003);
{\bf C38} 309 (2004).

\bibitem{KKSS}B.A.~Kniehl, G.~Kramer, I.~Schienbein and H.~Spiesberger,
Phys.\ Rev.\ {\bf D71} 014018 (2005);
Eur.\ Phys.\ J.\ {\bf C41} 199 (2005).

\bibitem{PRL}B.A.~Kniehl, G.~Kramer, I.~Schienbein and H.~Spiesberger,
Phys.\ Rev.\ Lett.\ {\bf 96} 012001 (2006).

\bibitem{Kneesch:2007ey}T.~Kneesch, B.A.~Kniehl, G.~Kramer and I.~Schienbein,
Nucl.\ Phys.\ {\bf B799} 34 (2008).

\bibitem{Kniehl:2008zz}B.A.~Kniehl, G.~Kramer, I.~Schienbein and
H.~Spiesberger,
Phys.\ Rev.\ {\bf D77} 014011 (2008).

\bibitem{ddata}G.~Alexander {\it et al.}\ (OPAL Collaboration),
Z.\ Phys.\  {\bf C72} 1 (1996);
K.~Ackerstaff {\it et al.}\ (OPAL Collaboration),
Eur.\ Phys.\ J.\ {\bf C1} 439 (1998);
R.~Barate {\it et al.}\ (ALEPH Collaboration),
Eur.\ Phys.\ J.\ {\bf C16} 597 (2000);
M.~Artuso {\it et al.}\ (CLEO Collaboration),
Phys.\ Rev.\ {\bf D70} 112001 (2004);
R.~Seuster {\it et al.}\ (Belle Collaboration),
Phys.\ Rev.\ {\bf D73} 032002 (2006).

\bibitem{Acosta:2003ax}D.~Acosta {\it et al.}\ (CDF Collaboration),
Phys.\ Rev.\ Lett.\ {\bf 91} 241804 (2003).

\bibitem{Bowler:1981sb}M.G.~Bowler,
Z.\ Phys.\ {\bf C11} (1981) 169.

\bibitem{BKK}J.~Binnewies, B.A.~Kniehl and G.~Kramer,
Phys.\ Rev.\ {\bf D58} 034016 (1998).

\bibitem{CDF}F.~Abe {\it et al.}\ (CDF Collaboration),
Phys.\ Rev.\ Lett.\ {\bf 75} 1451 (1995);
D.~Acosta {\it et al.}\ (CDF Collaboration),
Phys.\ Rev.\ {\bf D65} 052005 (2002).

\bibitem{Cacciari:2003uh}M.~Cacciari, S.~Frixione, M.L.~Mangano, P.~Nason and
G.~Ridolfi,
JHEP {\bf 0407} 033 (2004).

\bibitem{bdata}K.~Abe {\it et al.}\ (SLD Collaboration),
Phys.\ Rev.\ Lett.\ {\bf 84} 4300 (2000);
Phys.\ Rev.\ {\bf D65} 092006 (2002); {\bf D66} 079905(E) (2002);
A. Heister {\it et al.}\ (ALEPH Collaboration),
Phys.\ Lett.\ {\bf B512} 30 (2001);
G. Abbiendi {\it et al.}\ (OPAL Collaboration),
Eur.\ Phys.\ J.\ {\bf C29} 463 (2003).

\bibitem{Kartvelishvili:1985ac}V.G.~Kartvelishvili and A.K.~Likhoded,
Yad.\ Fiz.\ {\bf 42} 1306 (1985)
[Sov.\ J.\ Nucl.\ Phys.\ {\bf 42} 823 (1985)].

\bibitem{Cacciari:2002xb}M.~Cacciari and E.~Gardi,
Nucl.\ Phys.\ {\bf B664} 299 (2003).

\bibitem{OPAL}G.~Alexander {\it et al.}\ (OPAL Collaboration),
Phys.\ Lett.\ {\bf B364} 93 (1995).

\bibitem{FONLL}M.~Cacciari, M.~Greco and P.~Nason,
JHEP {\bf 9805} 007 (1998);
M.~Cacciari and P.~Nason,
Phys.\ Rev.\ Lett.\ {\bf 89} 122003 (2002).

\bibitem{CDF1}D.~Acosta {\it et al.}\ (CDF Collaboration),
Phys.\ Rev.\ {\bf D71} 032001 (2005).
  
\bibitem{CDF2}A.~Abulencia {\it et al.}\ (CDF Collaboration),
Phys.\ Rev.\  {\bf D75} 012010 (2007).

\bibitem{Martin:1992as}A.D.~Martin, W.J.~Stirling and R.G.~Roberts,
Phys.\ Rev.\ {\bf D47} 867 (1993).

\bibitem{pdg}W.M. Yao {\it et al.}\ (Particle Data Group),
J. Phys.\ {\bf G33} 1 (2006).

\bibitem{Happacher}F.~Happacher, P.~Giromini and F.~Ptohos,
Phys.\ Rev.\ {\bf D73} 014026 (2006).

\bibitem{CDF3} J.A. Kraus, Ph.D. thesis, University of Illinois,
Urbana-Champaign, 2006, Report No.\ FERMILAB-THESIS-2006-47;
A.~Annovi (CDF Collaboration),
J.\ Phys.: Conf.\ Ser.\ {\bf 110} 022003 (2008).

\end{thebibliography}
\end{document}